\documentclass[12pt, 3p]{elsarticle}
\usepackage{amssymb}
\usepackage{amsmath}
\usepackage{amsfonts}
\usepackage{setspace}
\usepackage{algorithm}
\usepackage{algorithmic}
\usepackage{graphicx}

\begin{document}

\begin{frontmatter}
\title{Approximate Aggregate Utility Maximization in Multi-Hop Wireless Networks using Distributed Greedy Scheduling}

\author[cedt]{Albert Sunny}
\ead{salbert@cedt.iisc.ernet.in}
\author[cedt]{Joy Kuri \corref{cor1}}
\ead{kuri@cedt.iisc.ernet.in}
\author[crl]{Nachiket Sahasrabudhe}
\ead{nachiket.sahasrabudhe@crlindia.com}

\cortext[cor1]{Corresponding author}
\address[cedt]{Centre for Electronics Design and Technology, \\
Indian Institute of Science, Bangalore-560012, India}
\address[crl]{Computational Research Laboratories, \\
Pune-411030, India}

\begin{abstract}
In this paper, we study the performance of greedy scheduling in multihop wireless networks, where the objective is aggregate utility maximization. Following standard approaches, we consider the dual of the original optimization problem. We note that the dual can be solved optimally, only with the knowledge of the maximal independent sets in the network. But computation of maximal independent sets is known to be NP-hard. Motivated by this, we propose a distributed greedy heuristic to address the problem of link scheduling. We evaluate the effect of the distributed greedy heuristic on aggregate utility maximization in detail, for the case of an arbitrary graph. We provide some insights into the factors affecting aggregate utility maximization in a network, by providing bounds on the maximum aggregate utility. We give simulation results for the approximate aggregate utility maximization achieved under distributed implementation of the greedy heuristic and find them close to the maximum aggregate utility obtained using optimal scheduling.
\end{abstract}

\begin{keyword}
cross-layer optimization, distributed scheduling, $\epsilon$-subgradient, greedy scheduling
\end{keyword}

\end{frontmatter}

\newdefinition{definition}{Definition}
\newproof{proof}{Proof}
\newtheorem{lemma}{\textit{Lemma}}
\newtheorem{corollary}{\textit{Corollary}}
\newtheorem{theorem}{\textit{Theorem}}
\newtheorem{proposition}{\textit{Proposition}}

\section{Introduction and Related Work}
We consider a wireless multihop mesh network in which some nodes are sources of traffic. Each source node is assumed to have an infinite backlog of traffic that is to be sent to a destination node. Further, each source is equipped with a utility function that is concave increasing in the average data rate that it can push through the network. We are interested in obtaining a distributed cross-layer scheme for joint congestion control, routing and link scheduling, such that the aggregate utility is maximized. 

Aggregate utility maximization seeks to strike a balance between high total throughput and fairness. At one extreme, a solution may maximize the sum of throughputs. However, in such a solution, some sources may see very low throughputs even though the sum of throughputs is maximized. At the other extreme, a solution may be max-min fair, in which the smallest throughput is as large as possible, and, among such solutions, the second smallest is as large as possible, and so on. But, such a solution may lead to a low value of aggregate throughput. The aggregate utility maximization approach lies in between these extremes.

Research into scheduling, routing and congestion control is several decades old, but has seen a lot of activity following the seminal paper of Tassiulas and Ephremides. The literature can be classified into two broad groups. In the first group, traffic arrives into the network according to some specified random processes that cannot be controlled and the objective is to find the ``capacity region'' i.e., the largest set of arrival vectors for which a scheduling and routing policy can be found ensuring stable operation \cite{tassi, stolyar, buche, shakko, shakko1, andrews}.

In the second group, each source equipped with a utility function has an infinite backlog of data to send and the objective is to maximize aggregate utility. Following the important paper of Kelly, Maulloo and Tan \cite{kelly}, researchers have addressed the issue of obtaining distributed controls to achieve the objective. The basic idea is that the network will provide congestion signals to the sources (in the form of ``prices''), and the sources will modify their data rates accordingly. The original problem is shown to decompose
into several subproblems, viz., congestion control, routing and scheduling, and the objective is to find distributed solutions to each \cite{doyle, doyle1, lopresti, lin, shroff}. We consider a problem that belongs to the second group above.

This programme was later carried out for wireless networks, where the additional aspect of wireless link scheduling appeared \cite{stolyar, bui, doyle1, tan, lin, ery1, gupta, ery}. 

The main results of this paper can be summarized as follows:
\begin{itemize}
\item We develop a distributed greedy heuristic for the scheduling problem under the $K$-hop link interference model.
\item We show that the distributed greedy heuristic leads to an $\epsilon$-subgradient which can be used to solve the problem of aggregate utility maximization.
\item Further, we evaluate the effect of the sub-optimal greedy schedule by looking at the convergence properties of the $\epsilon$-subgradient method.
\item We also provide some insights into the factors affecting aggregate utility maximization in a network, by providing bounds on the maximum aggregate utility.
\end{itemize}

\section{System Model and Mathematical Formulation}
\label{section:model}
\subsection{Primal optimization problem}

We assume the network to be a directed graph $\mathcal{G} = (\mathcal{N}, \mathcal{L})$, where $\mathcal{N}$ represents the set of nodes in the network and $\mathcal{L}$ represents the set of wireless links in the network. As in [1], we assume that a bidirectional wireless link $(i, j) \in \mathcal{L}$ exists in the network, if nodes $i$ and $j$ are within transmission range of each other. We assume that each link $l \in \mathcal{L}$ has a capacity denoted by $C_l$. Let
$$ C = \max_{l \in \mathcal{L}}C_l$$
i.e., $C$ represents the maximum link capacity in the network. Now, we can represent the capacity of a link $l \in \mathcal{L}$ as
$$ C_l = \alpha_l \cdot C \quad \textrm{where} \quad 0 < \alpha_l \leq 1$$
The $\alpha_l$'s are dimensionless and we formulate the problem in terms of these dimensionless quantities. Let $\mathcal{F}$ denote the set of all end-to-end multihop flows present in the network. For each flow $f \in \mathcal{F}$, $s(f)$ and $d(f)$ represent the source and destination nodes of the flow $f$ respectively. We assume that the source nodes have an infinite backlog of data. Let $x_f$ denote the  data rate associated with the flow $f$ and $y_{fl}$ denote the part of the flow $f$ that is carried by the link $l$. Let $\mathbf{y}_f = (y_{fl}), \, l \in \mathcal{L}$ be a vector representing the part of flow $f$ carried by each link in the network. 

In this paper, we model the inter-link interference using the $K$-hop link interference model. Here, we reproduce some definitions from \cite{sharma} in order to define the interference model.

\begin{definition}
Let $d_S(x, y)$ denote the shortest distance (in terms of number of links) between nodes $x, y \in \mathcal{N}$. Define a function $d : (\mathcal{L}, \mathcal{L}) \to \mathbb{N}$ as follows: For links $l_u = (u_1, u_2), \, l_v = (v_1, v_2) \in \mathcal{L}$, let
$$ d(l_u, l_v) = \min_{i,j \in \{1,2 \}} d_S(u_i, v_j)$$
\end{definition}

In the $K$-hop link interference model, we assume that any two links $l_1$ and $l_2$ for which $d(l_1, l_2) < K$, will interfere with each other and hence cannot be active simultaneously.

A \emph{maximal independent set} of links ($\mathcal{I}$), is a set of links in which no two links of the set $\mathcal{I}$ interfere with each other under the given interference model and no other link can be added to the set $\mathcal{I}$ without violating the interference constraints. We represent a maximal independent set of links $\mathcal{I}$ by a column vector $\mathbf{r}_{\mathcal{I}}$ of size $|\mathcal{L}|$. If a link $l \in \mathcal{I}, \textrm{then } \mathbf{r}_\mathcal{I}(l) \textrm{ is } \alpha_l; \textrm{ else it is } 0$. We represent the collection of all the maximal independent sets by the matrix $\mathbf{M}$, columns of which are the $\mathbf{r}_{\mathcal{I}} s$. Let there be $J$ maximal independent sets present in the network. We represent a schedule associated with them by a $J$ sized column vector $\mathbf{a}$, where the $i^{\textrm{th}}$ entry represents the fraction of time the independent set represented the $i^{\textrm{th}}$ column of the matrix $\mathbf{M}$ is active. We associate a strictly concave, twice differentiable, increasing utility function $U(x_f)$ with every end-to-end flow $f \in \mathcal{F}$.  We also assume $U(0)$ to be less than or equal to zero. Such an assumption is natural, because it asserts that, if a source is not able to sustain a positive rate of data transfer, then the ``utility'' for that source is a finite non-positive value.

Let $\mathbf{A}$ be the $|\mathcal{N}| \times |\mathcal{L}|$ node-link incident matrix. Then, the formal representation of the problem is as follows:
\begin{eqnarray}
\max_{ x_f \geq 0, \mathbf{a} \geq 0} && \sum_{f \in \mathcal{F}}{U(x_f)} \nonumber \\
\textrm{Subject to : } \mathbf{A} \mathbf{y}_f &=& \mathbf{u}_f \, , \, \forall f \in \mathcal{F} \\
x_f &\leq& 1, \quad \forall f \in \mathcal{F} \\
\sum_{f \in \mathcal{F}} \mathbf{y}_f &\leq& \mathbf{M} \mathbf{a} \label{eq:flow} \\
\sum^{J}_{k=1} a_k &=& 1 
\end{eqnarray}
where $\mathbf{u}_f$ represents a $N$-sized column vector such that $u_f(s(f)) = x_f$, $u_f(d(f)) = -x_f$ and all other entries are zero. Here, the first constraint ensures flow conservation at every node in the network. The second constraint ensures that the data rates of the flows are less than or equal to the maximum link capacity in the network. The next constraint, represented by Equation \eqref{eq:flow} ensures that the aggregate flow on each link is less than or equal to the effective capacity of the link. The fourth constraint ensures that there are no idle slots in the schedule.  

\subsection{Dual problem}
The primal optimization problem is a convex optimization problem with affine constraints. By applying Slater's condition \cite{slater}, it can be shown that this problem has no duality gap. To obtain a solution in a distributed manner, we consider the dual problem as in \cite{kelly, doyle}. The capacity constraints given by Equation \eqref{eq:flow} are relaxed to obtain the Lagrange variables. These Lagrange variables behave as link prices and we represent them by a vector $\mathbf{p}$. Now, the dual problem associated with the primal problem can be stated as follows:
\begin{eqnarray}
\min_{\mathbf{p} \geq 0} D(\mathbf{p}) \nonumber
\end{eqnarray}
where
\begin{eqnarray}
D(\mathbf{p}) = && \max_{x_f \geq 0, \mathbf{a} \geq 0} \left( \sum_{f \in \mathcal{F}} \left( U(x_f)- \mathbf{p}^T \left( \mathbf{y}_f - \mathbf{M a} \right) \right) \right) \label{eq:dual}
\end{eqnarray}
\begin{eqnarray*}
\mathrm{Subject\ to:\ } \mathbf{Ay}_f &=& \mathbf{u}_f, \forall f \in \mathcal{F}  \\
x_f &\leq& 1, \quad \forall f \in \mathcal{F} \\
\sum_{k=1}^J a_k &=& 1
\end{eqnarray*}
We note that given a price vector $\mathbf{p}$, the RHS of Equation \eqref{eq:dual} can be written as a sum of two functions of price vector $\mathbf{p}$ i.e.,
$$D(\mathbf{p}) = D_1(\mathbf{p}) + D_2(\mathbf{p})$$
Here, $D_1(\mathbf{p})$ corresponds to the \emph{congestion control and routing problem},  whereas $D_2(\mathbf{p})$ represents the \emph{link scheduling problem}. 

\section{Congestion Control and Routing Subproblem}
\label{section:congestion}
\begin{eqnarray*}
D_1(\mathbf{p}) = && \max_{x_f \geq 0}   \sum_{f \in \mathcal{F}} \left( U(x_f) - \sum^{|\mathcal{L}|}_{i=1} p_iy_{fi} \right)  \\
\textrm{Subject to: } && \mathbf{A} \mathbf{y}_f = \mathbf{u}_f \, , \, \forall f \in \mathcal{F} \\
&& x_f \leq 1, \quad \forall f \in \mathcal{F} 
\end{eqnarray*}

For a given vector of link prices $\mathbf{p}$, each source solves the problem of how much traffic to send into the network so as to maximize its net utility. Since the maximum value that $x_f$ can attain is $1$, we can obtain the optimal value of $x_f$ as
\begin{eqnarray} \label{eq:xf}
x_f = \min \{U^{'^{-1}} (p(f)), 1 \}
\end{eqnarray}
where $p(f)$ is cost of the least-priced path between $s(f)$ and $d(f)$ for a given $\mathbf{p}$. Since all the link capacities are normalized with respect to $C$, the $x_f$ obtained from Equation \ref{eq:xf} represents the data rate normalized with respect to $C$. Hence, the actual rate at which traffic is to be injected into the network is given by $x_f \cdot C$. The least-priced path for a flow $f$ can be found by using a modified \emph{Distributed Bellman-Ford} algorithm, which uses the \emph{link price} as the metric in place of hop count. If there are multiple least-priced paths, then the traffic can be split among them in any arbitrary manner. 

\section{Scheduling Subproblem}
\label{section:scheduling}
\begin{eqnarray*}
D_2(\mathbf{p}) &=&  \max_{\mathbf{a} \geq 0}  \mathbf{p}^{t} \mathbf{c}  \\
\textrm{Subject to: } \mathbf{c} &=& \sum_{I} \mathbf{r}_{I} a_{I} \quad \textrm{and}  \quad \sum_{I} a_I = 1 
\end{eqnarray*}
It can be shown that for a given vector of link prices $\mathbf{p}$, the solution to this problem is to schedule an independent set of ``maximum aggregate capacity-weighted price'' \cite{slater} i.e., a maximal independent set of links $\mathcal{I}_{opt}$ is the optimal solution if 
$$ \mathcal{I}_{opt} = \arg \max_{\mathcal{I}} \left( \sum_{l \in \mathcal{I}} \alpha_l p_l \right) $$
Thus, an optimal scheduling vector $\mathbf{a}_{opt}(\mathbf{p})$ has all its entries zero, except the one corresponding to an independent set with maximum aggregate ``\emph{capacity-weighted price}''.  Computing the optimal schedule requires knowledge of all the maximal independent sets in the network and is known to be NP-hard \cite{sharma}. Hence, it is difficult to realize the same in a distributed manner for an adhoc network. We employ a distributed greedy heuristic to obtain a distributed solution for the links scheduling problem. We denote the schedule obtained using the distributed greedy heuristic by $\mathbf{a}_{dgrd}(\mathbf{p})$. 

\subsection{Distributed Greedy Heuristic}
In this section, we consider a centralized greedy heuristic and examine the possibility of implementing the same in a distributed manner. Here, we present the Greedy Heuristic.

\begin{algorithm}[H]
\caption{Centralized Greedy Heuristic}
\begin{algorithmic}[1]
\STATE Set $W:= \phi$ and $i := 1$.
\STATE Arrange the links of $\mathcal{L}$ in descending order of \emph{capacity-weighed} price (i.e., $\alpha_l p_l)$, starting with $l_1, l_2,...  .$ 
\STATE If $W \cup l_i$ is a valid $K$-matching, then, $W := W \cup l_i$ , $i = i + 1$.
\STATE Repeat Step 3 for all links in $\mathcal{L}$.
\end{algorithmic}
\end{algorithm}
\noindent
Here, a set of edges $W$ is a valid $K$-matching if $\forall l_1, l_2 \in W$ with $l_1 \neq l_2$, we have $d(l_1, l_2) \geq K$.

In \cite{joo1}, the authors have shown that the Greedy Maximal Schedule (GMS) achieves the full capacity region in tree networks under the $K$-hop interference model and  that the worst-case efficiency ratio of GMS in geometric unit-disc graphs is between $\frac{1}{6}$ and $\frac{1}{3}$. 

Next, we present a distributed version of the greedy heuristic. The algorithm for the distributed greedy heuristic is described below and is implemented at every node $n \in \mathcal{N}$.

\begin{figure}[h]
\centering
\includegraphics[width=\linewidth, height=0.94in]{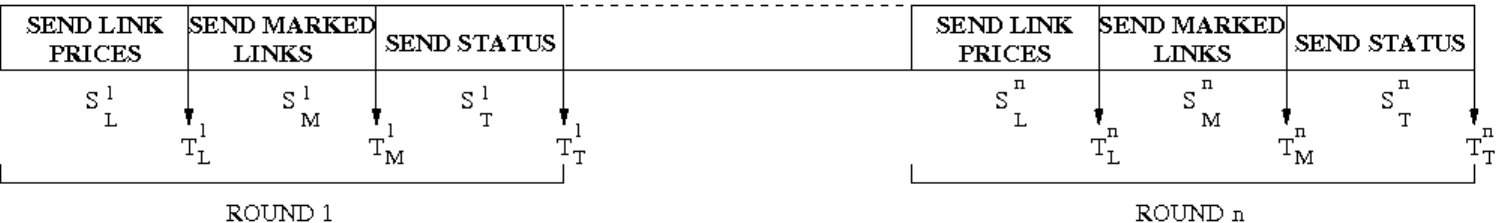}
\caption {Slot division of the distributed greedy algorithm.}
\label{fig:ch3_dp}
\end{figure}

The algorithm is described in terms of the messages exchanged between nodes in each slot of the $m^{\mathrm{th}}$ \textit{ROUND} and the decisions that are made at the slot boundaries after the completion of the given slot. Together, the \textit{SEND LINK PRICES} slot, the \textit{SEND MARKED LINKS} slot and the \textit{SEND STATUS} slot constitute a \textit{ROUND} as shown in Figure \ref{fig:ch3_dp}.

Each link $l$ can have four possible states namely \textit{OPEN (O)}, \textit{CHECK (CH)}, \textit{MARKED (M)} and \textit{CLOSED (CL)}. A link $l \in \mathcal{L}$ is called an ``attached link'' of node $n \in \mathcal{N}$, if node $n$ is an end point of the link $l$ and the link $l$ is directed outwards from node $n$. Initially, all links are set to \textit{OPEN}, the algorithm status set is to \textit{DO NOT TERMINATE} and  $m$ is set as 1.

\vspace{2mm}
\begin{spacing}{0.8}
\hrule
\vspace{2mm}
\noindent
\textbf{Algorithm 2} Pseudo-code for Distributed Greedy Heuristic \\
\hrule
\end{spacing}
\begin{algorithmic}[h]
\STATE \textit{\textbf{In slot $S^m_L$}}
\end{algorithmic}
\begin{algorithmic}[1]
\STATE Find the highest \emph{capacity-weighed} priced (i.e., $\alpha_l p_l$) link among all the \textit{OPEN} attached links and disseminate this information over  the $(K+1)$-hop neighbourhood.
\end{algorithmic}
\begin{algorithmic}
\STATE \textit{\textbf{At time $T^m_L$}}
\end{algorithmic}
\begin{algorithmic}[1]
\IF{at least one attached link is \textit{OPEN}}
	\STATE sort the \textit{OPEN} attached  links in descending order of their respective \emph{capacity-weighed} price. Let $l^{'}_{max}$ be the maximum \emph{capacity-weighed} priced link among the attached \textit{OPEN} links.
	\IF{no \emph{capacity-weighed} prices are received from the (K+1)-hop neighbours}
		\STATE link $l^{'}_{max}$ is \textit{MARKED} and all other \textit{OPEN} attached links are \\
		\textit{CLOSED} and go to 17.
	\ELSE
		\STATE sort the received \textit{OPEN} links in descending order of their respective \emph{capacity-weighed} price. Let $l_{max}$ be the maximum \emph{capacity-weighed} priced link among the received \textit{OPEN} links. 
	\ENDIF
	\IF{$(p_{l^{'}_{max}} > p_{l_{max}} )$}
		\STATE link $l^{'}_{max}$ is \textit{MARKED} and all other \textit{OPEN} attached links are \\
		\textit{CLOSED}.
	\ELSE
		\FORALL{\textit{OPEN} attached link $l$,}
			\IF{$(d(l, l_{max}) < K) $} 
			\STATE link $l$ is set to \textit{CHECK}.
			\ENDIF
		\ENDFOR
	\ENDIF
\ENDIF	
\end{algorithmic}
\hrule
\vspace{2mm}

The links that are \textit{MARKED} are the maximum \emph{capacity-weighed} priced links in their corresponding $(K+1)$ hop neighbourhoods. Also, the links that are moved to the \textit{CLOSED} state are certain to have a \textit{MARKED} link within $K$ hop link distance. If a link moves to \textit{MARKED} or \textit{CLOSED} state in \textit{ROUND} $m$, it will continue to remain in that state until the  algorithm terminates (\textit{i.e.,} these states are absorbing states). These links will not participate in price dissemination in the subsequent \textit{ROUND}s. Links that are in the \textit{MARKED} state will be scheduled, upon termination of the algorithm. The reason for introducing a \textit{CHECK} state is to help resolve ambiguities that arise due to decisions based on information from local neighbourhood.  

\addtocounter{algorithm}{-1}

\vspace{2mm}
\begin{spacing}{0.8}
\hrule
\vspace{1mm}
\noindent
\textbf{Algorithm 2} Pseudo-code for Distributed Greedy Heuristic \\
\hrule
\end{spacing}
\begin{algorithmic}[h]
\STATE \textit{\textbf{In slot $S^m_M$}}
\end{algorithmic}
\begin{algorithmic}[1]
\IF{any of the attached links is \textit{MARKED}}
	\STATE disseminate this information to the $(K+1)$-hop neighbourhood. 
\ENDIF
\end{algorithmic}
\begin{algorithmic}
\STATE \textit{\textbf{At time $T^m_M$}}
\end{algorithmic}
\begin{algorithmic}[1]
\FOR{each attached link $l$ in state \textit{CHECK},}
	\IF{$(d(l,\textrm{received \textit{MARKED} link})< K))$ for at least one received \\
	\textit{MARKED} link}
		\STATE link $l$ is \textit{CLOSED}.
	\ELSE
		\STATE link $l$ remains in \textit{CHECK} state. 
	\ENDIF
\ENDFOR
\STATE	\textit{OPEN} the highest \emph{capacity-weighed priced} attached \textit{CHECK} link.
\STATE Algorithm status is set to \textit{TERMINATE} at nodes which have no \textit{OPEN} or \textit{CHECK} links.
\end{algorithmic}
\hrule
\vspace{2mm}

During the price dissemination slot, links will move into \textit{CHECK} state if they see a higher-\emph{capacity-weighed} priced interfering link, but are unable to decide if such a link will get \textit{MARKED}. In the \textit{SEND MARKED LINKS} slot, \textit{CHECK} links get to know if there is indeed a higher \emph{capacity-weighed} priced \textit{MARKED} link interfering with it. If so, the link in \textit{CHECK} state is \textit{CLOSED}.

\addtocounter{algorithm}{-1}

\vspace{2mm}
\begin{spacing}{0.8}
\hrule
\vspace{1mm}
\noindent
\textbf{Algorithm 2} Pseudo-code for Distributed Greedy Heuristic \\
\hrule
\end{spacing}
\begin{algorithmic}[H]
\STATE \textit{\textbf{In slot $S^m_T$}}
\end{algorithmic}
\begin{algorithmic}[1]
\IF{at least one attached link is \textit{OPEN} or in \textit{CHECK}}
	\STATE send a \textit{DO NOT TERMINATE} message to all nodes in the $(K+1)$-hop neighbourhood.
\ELSIF{a \textit{DO NOT TERMINATE} message is received}
    \STATE send a \textit{DO NOT TERMINATE} message to all nodes in the $(K+1)$ 
    \\hop neighbourhood.
\ENDIF
\end{algorithmic}
\begin{algorithmic}
\STATE \textit{\textbf{At time $T^m_T$}}
\end{algorithmic}
\begin{algorithmic}[1]
\IF{no \textit{DO NOT TERMINATE} message is received}
	\STATE the algorithm has terminated, schedule all \textit{MARKED} links.
\ELSE
	\STATE go to the $(m+1)^{\mathrm{th}}$ \textit{ROUND}.
\ENDIF 
\end{algorithmic}
\hrule

The local termination condition is that no attached link is in \textit{OPEN} or \textit{CHECK} state. In the above slot, this information is conveyed to all the other nodes in the network in a distributed manner. This makes sure that the algorithm terminates in a synchronous fashion at each node.
 
\begin{figure}[h]
\centering
\includegraphics[width=\linewidth]{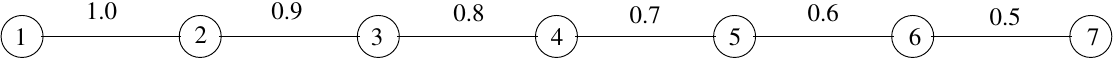}
\caption{An example to illustrate the distributed greedy scheduling algorithm.}
\label{fig:example1}
\end{figure}

Next, we illustrate the working of our distributed greedy heuristic under a $2$-hop link interference model with a couple of examples. First, let us consider a linear network with 7 nodes as shown in Figure \ref{fig:example1}. For data transfer, only links (1,2), (2,3), (3,4), (4,5), (5,6) and (6,7) are considered; but control traffic can flow in the opposite direction too. In Table \ref{table:example1_table}, we show the link states against the time when different decisions are made.

\begin{table}[h]
\begin{center}
\begin{tabular}{|c|c|c|c|c|c|c|}
\hline T & (1,2)  & (2,3) & (3,4) & (4,5) & (5,6) & (6,7)  \\ 
\hline $0$ & \textit{O} & \textit{O} & \textit{O} & \textit{O} & \textit{O} & \textit{O} \\ 
\hline $T^1_L$ & \textit{M} & \textit{CH} & \textit{CH} & \textit{CH} & \textit{CH} & \textit{CH} \\ 
\hline $T^1_M$ & \textit{M} & \textit{CL} & \textit{CL} & \textit{O} & \textit{O} & \textit{O} \\ 
\hline $T^2_L$ & \textit{M} & \textit{CL} & \textit{CL} & \textit{M} & \textit{CH} & \textit{CH} \\ 
\hline $T^2_M$ & \textit{M} & \textit{CL} & \textit{CL} & \textit{M} & \textit{CL} & \textit{CL} \\ 
\hline 
\end{tabular}
\end{center} 
\caption{Table showing the states of the links in Figure \ref{fig:example1} against the various slots of the distributed greedy scheduling algorithm.}
\label{table:example1_table}
\end{table}

\begin{figure}[h]
\centering
\includegraphics[width=\linewidth]{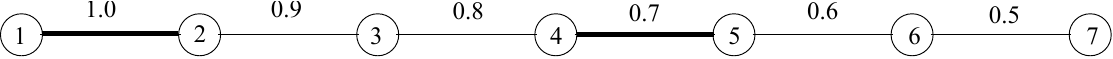}
\caption{Figure showing scheduled links (in bold) using distributed greedy scheduling on the example in Figure \ref{fig:example1}.}
\label{fig:example11}
\end{figure}

In the first \textit{ROUND}, only link (1,2) is \textit{MARKED}. All other links see a higher \emph{capacity-weighed} priced interfering link and thus move into the \textit{CHECK} state. Then, link (1,2) announces it is \textit{MARKED}. Upon reception of this information, links (2,3) and (3,4) are \textit{CLOSED}, since they interfere with link (1,2). But all other links are moved to \textit{OPEN}, since they do not find any interfering \textit{MARKED} link. This process repeats until the network has no \textit{OPEN} or \textit{CHECK} links. The set of links that are scheduled after the algorithm terminates, are shown in bold, in Figure \ref{fig:example11}.

\begin{figure}[h]
\centering
\includegraphics[width=\linewidth]{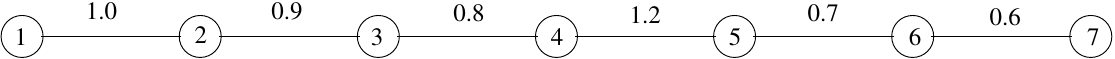}
\caption{Another example to illustrate the distributed greedy scheduling algorithm.}
\label{fig:example2}
\end{figure}

\noindent
Next, we consider a 7 node linear network shown in Figure \ref{fig:example2}.

As in the previous example, we show the link states against the time when different decisions are made in Table \ref{table:example2_table} 

\begin{table}[h]
\begin{center}
\begin{tabular}{|c|c|c|c|c|c|c|}
\hline T & (1,2)  & (2,3) & (3,4) & (4,5) & (5,6) & (6,7)  \\
\hline $0$ & \textit{O} & \textit{O} & \textit{O} & \textit{O} & \textit{O} & \textit{O} \\ 
\hline $T^1_L$ & \textit{O} & \textit{CH} & \textit{CH} & \textit{M} & \textit{CH} & \textit{CH} \\ 
\hline $T^1_M$ & \textit{O} & \textit{CL} & \textit{CL} & \textit{M} & \textit{CL} & \textit{CL} \\ 
\hline $T^2_L$ & \textit{M} & \textit{CL} & \textit{CL} & \textit{M} & \textit{CL} & \textit{CL} \\ 
\hline $T^2_M$ & \textit{M} & \textit{CL} & \textit{CL} & \textit{M} & \textit{CL} & \textit{CL} \\ 
\hline 
\end{tabular}
\end{center} 
\caption{Table showing the states of the links in Figure \ref{fig:example2} against the various slots of the distributed greedy scheduling algorithm.}
\label{table:example2_table}
\end{table}

\begin{figure}[H]
\centering
\includegraphics[width=\linewidth]{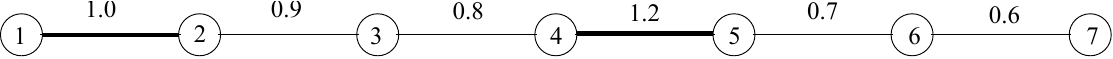}
\caption{Figure showing selected links (in bold) using distributed greedy scheduling on the example in Figure \ref{fig:example2}.}
\label{fig:example21}
\end{figure}

In this example, the highest \emph{capacity-weighed} priced link is located in the middle of the network. As a result, more links are moved into the \textit{CLOSED} state after the first \textit{ROUND}. 
The only remaining \textit{OPEN} link, \textit{i.e.,} link (1,2), is \textit{MARKED} in the subsequent 
\textit{ROUND}. The set of links that are scheduled after the algorithm terminates, are shown in bold, in Figure \ref{fig:example21}.

One can compute the centralized greedy schedule for the above examples easily and verify that the distributed greedy schedule matches 
it. 

\subsection{Performance of Greedy Scheduling}

In this section, we show analytically that the distributed greedy heuristic schedules the same set of links as the centralized greedy heuristic. These results are from our previous work \cite{albert}, but are included here for completeness. 
\begin{lemma} \label{lemma:finite_time}
The algorithm terminates in finite time.
\end{lemma}
\begin{proof}
In Appendix A.
\end{proof}
Let us assume that no two links have equal \emph{capacity-weighted} prices i.e., $\forall i,j \in \mathcal{L}, p_i \alpha_i \neq p_j \alpha_j$; or in other words, we have a unique way of breaking tie among equal \emph{capacity-weighted} priced links.
\begin{lemma} \label{lemma:ft_lemma2}
For every link $i$ \textit{CLOSED} before \textit{ROUND} $m$, there must be at least one link \textit{MARKED} before \textit{ROUND} $m$, that interferes with it.
\end{lemma}
\begin{proof}
In Appendix A.
\end{proof}
The additional \textit{CHECK} state helps to resolve ambiguities that arise in deciding the highest \emph{capacity-weighted} priced link based on information from a local neighbourhood. If a link $i$ sees an interfering higher \emph{capacity-weighted priced} link in its neighbourhood, it will move into \textit{CHECK} state. Such a link, will subsequently move to the \textit{CLOSED} state, if it receives an announcement to \textit{CLOSE} from an interfering higher \emph{capacity-weighted} priced \textit{MARKED} link. Now, we would like to formally show that the introduction of \textit{CHECK} state does not alter the performance of the algorithm.
\begin{lemma} \label{lemma:ft_lemma3}
At the beginning of a \textit{ROUND}, consider the globally highest capacity-weighted priced link among links that are neither \textit{CLOSED} nor \textit{MARKED}; such a link will not be in \textit{CHECK} state. Or, it can be restated as; at the end of a round, the highest capacity-weighted priced \textit{CHECK} link is \textit{OPEN}ed.
\end{lemma}
\begin{proof}
In Appendix A.
\end{proof}

\begin{theorem} \label{theorem:same_links}
The distributed greedy algorithm and the centralized greedy algorithm schedule the same links.
\end{theorem}
\begin{proof}
In Appendix A.
\end{proof}

Next, we quantify the deviation of our distributed greedy schedule from the optimal one.

\begin{definition} 
\cite{sharma}  The $K$-hop interference set of a link $l  \in \mathcal{L}$, denoted by $\mathcal{I}_K(l, \mathcal{L})$, is the set of links $m \in \mathcal{L}$ such that $d(l, m) < K$.
\end{definition} 

We call a subset $\mathcal{D}$ of $\mathcal{I}_K(l,\mathcal{L})$ ``\emph{K-maximal}'', if no link $m \in \mathcal{I}_K(l, \mathcal{L}) \setminus \mathcal{D}$ can be added to $\mathcal{D}$, without violating the interference constraints. In other words, for all $m \in \mathcal{I}_K(l, \mathcal{L})$ and all $n \in \mathcal{D}$, $d(m , n) < K$.

\begin{definition}
\cite{sharma} The $K$-hop interference degree of a link $l  \in \mathcal{L}$, denoted by $d_K(l)$, is defined as 
$$d_K(l) = \max_{ S \subseteq I_K(l, \mathcal{L}): \mathcal{D} \, is \, K-maximal } |\mathcal{D}|$$
\end{definition}

\begin{definition}
\cite{sharma}  The $K$-hop interference degree of a graph $G( \mathcal{N},\mathcal{L} )$, denoted by $d_{K} (G)$, is defined as
$$d_{K} (G) = \max_{l \in \mathcal{L}}  \, d_{K}(l)$$
\end{definition} 

For a given price vector $\mathbf{p}$, let $S_{opt}(\mathbf{p})$, $S_{grd}(\mathbf{p})$ and $S_{dgrd}(\mathbf{p})$ represent the aggregate \emph{capacity-weighted prices} of the independent sets corresponding to the optimal  schedule, greedy schedule  and distributed greedy schedule, respectively. From Theorem \ref{theorem:same_links}, we know that $S_{dgrd}(\mathbf{p}) = S_{grd}(\mathbf{p})$. Substituting this in a theorem from \cite{sharma}, we have
\begin{theorem} \label{theorem:sharma}
$$\frac{S_{opt}(\mathbf{p})}{S_{dgrd}(\mathbf{p})} \leq  d_{K}(G)$$
Further, there exists a graph $G$ for which the above ratio is exactly $d_{K}(G)$.
\end{theorem}
\begin{proof}
We refer to Theorem 5 in \cite{sharma}.
\end{proof}
\section{$\epsilon$-Subgradient Method}
\label{section:epsilon}
In order to solve the dual problem, we have to minimize the dual objective function $D(\mathbf{p})$. The natural strategy would be to consider the negative gradient and take a small step in that direction. However, we notice that $D(\mathbf{p})$ is not differentiable in $\mathbf{p}$ everywhere. This raises the possibility of using subgradients.

\begin{lemma} \label{lemma:convex}
$D(\mathbf{p})$ is a convex function of $\mathbf{p}$.
\end{lemma}
\begin{proof}
In Appendix B.
\end{proof}

\begin{definition} \label{defn:subgradient}
\cite{bert} Given a convex function $D(\mathbf{p}) : \mathcal{R}^n \to \mathcal{R}$, we say that a vector $\mathbf{h}(\mathbf{p}) \in \mathcal{R}^n$ is a $\epsilon$-subgradient of $D(\mathbf{p})$ at point $\mathbf{p} \in \mathcal{R}^n$, if $D(\overline{\mathbf{p}}) \geq D(\mathbf{p}) - \epsilon + (\overline{\mathbf{p}} - \mathbf{p})^T \mathbf{h}(\mathbf{p}), \, \,  \forall \, \overline{\mathbf{p}} \in \mathcal{R}^n$.
\end{definition} 

\begin{proposition}  \label{prop:subgradient}
For the given price vector $\mathbf{p}$, the price updation rule leads to an $\epsilon(\mathbf{p})$-subgradient $\mathbf{h}(\mathbf{p})$ associated with price vector $\mathbf{p}$, given by 
$$\mathbf{h}(\mathbf{p}) = \mathbf{c}_{dgrd}(\mathbf{p}) - \mathbf{y}(\mathbf{p}) \quad \textrm{and} \quad \epsilon(\mathbf{p})=\mathbf{p}^{T}(\mathbf{c}_{opt}(\mathbf{p}) - \mathbf{c}_{dgrd}(\mathbf{p}))$$
where the $l^{\mathrm{th}}$ entry of $\mathbf{c}_{dgrd}(\mathbf{p})$ is $\alpha_l$ if link $l$ is present in the distributed greedy schedule $\mathbf{a}_{dgrd}(\mathbf{p})$ associated with the price vector $\mathbf{p}$, else it is zero; the 
$l^{\mathrm{th}}$ entry of $\mathbf{c}_{opt}(\mathbf{p})$ is $\alpha_l$ if link $l$ is present in the optimal schedule $\mathbf{a}_{opt}(\mathbf{p})$ associated with the price vector $\mathbf{p}$, else it is zero; further, the $l^{\mathrm{th}}$ entry of 
$\mathbf{y}(\mathbf{p})$ represents the aggregate flow on link $l \in \mathcal{L}$ when the price vector is $\mathbf{p}$.
\end{proposition} 
\begin{proof}
In Appendix B.
\end{proof}

The $\epsilon$-subgradient method can be used to generate a sequence of dual feasible points according to the iteration \cite{bert}
$$\mathbf{p}[j+1]  = \left( \mathbf{p}[j] -  \delta \cdot \mathbf{h}(\mathbf{p}[j]) \right)^{+}$$
Here, $(x)^{+} = \max (x, 0)$ and $\delta$ is the constant step size associated with the subgradient algorithm.
Hence, we update the price of link $l$ going from the $j^{\textrm{th}}$ to the $(j+1)^{\textrm{th}}$ iteration using the following equation:
\begin{eqnarray}
p_l[j + 1] = \left( p_l[j] +  \delta \left( y_l[j] - c_l[j] \right) \right)^{+}
\end{eqnarray}
where $y_l[j]$ represents the aggregate flow on link $l$ in the $j^{\textrm{th}}$ iteration obtained from the optimal solution of the routing and congestion control problem. If link $l$ is part of the distributed greedy schedule for the given price vector $p[j]$ in the $j^{\textrm{th}}$ iteration, then $c_l[j]$ takes the value $\alpha_l$, else it is zero.

Next, we provide a theorem that helps us establish a bound on the deviation of the approximately maximized aggregate utility from the maximum aggregate utility achievable. Since it has been shown in \cite{lopresti}, that the subgradient method with constant step size results in oscillations in ``steady-state'' (limit cycles), we only look at the convergence of the dual in the cesaro sense and not in the strict sense.

\begin{theorem} \label{th:convergence}
If $\lim_{j \to \infty} \epsilon(\mathbf{p}[j]) \leq \epsilon$ and $||\mathbf{h}(\mathbf{p}[j]) ||_2 \leq H < \infty \, \forall j$, then  	
\begin{eqnarray*}
D(\mathbf{p}^{*}) \leq \lim_{j \to \infty} \frac{1}{j} \sum^{j-1}_{\tau = 0 }D(\mathbf{p}[\tau]) \leq  D(\mathbf{p}^{*}) + \frac{\delta H^2}{2}  + \epsilon
\end{eqnarray*}
where $\mathbf{p}^{*}$ is an optimal solution to the dual problem $D(\mathbf{p})$. 
\end{theorem} 
\begin{proof}
In Appendix B.
\end{proof}

Theorem \ref{th:convergence} tells us that the average of the dual function obtained using the distributed greedy algorithm lies in a band above the optimal dual value.

\section{Conclusion and Future Work}
\label{section:conclude}

The scheduling problem is known to be a bottleneck in the cross-layer optimization approach. We have relax the optimality requirement and proposed a distributed greedy heuristic that schedules an independent set of links. Further, we have also quantified the effect of the sub-optimal distributed greedy heuristic by established. Even though we bound the performance of the distributed greedy scheduling in Theorem \ref{th:convergence}, it seems that one can obtain a tighter bounds atleast for specific type of networks. We would also like to address this problem in our future work.

\label{section:references}
\bibliographystyle{IEEEtran}
\bibliography{mainfile}

\appendix
\section*{Appendix A.}
\noindent
Let $\mathcal{L}^m_O$  be the set of \textit{OPEN} links before \textit{ROUND} $m$. \\
Let $\mathcal{L}^m_C$  be the set of \textit{CLOSED} links before \textit{ROUND} $m$. \\
Let $\mathcal{L}^m_H$  be the set of \textit{CHECK} links before \textit{ROUND} $m$. \\
Let $\mathcal{L}^m_M$  be the set of \textit{MARKED} links before \textit{ROUND} $m$. 
\subsection*{\textbf{Proof of Lemma \ref{lemma:finite_time}}}
\begin{proof}
If the set $\mathcal{L}^{m}_{O}$ is null, the algorithm terminates. Hence, we note that there is at least one \textit{OPEN} link at the beginning of each round. Let 
$$l^m = arg \, \max_{l \in \mathcal{L}^m_O} \, p_l \alpha_l$$
be the global maximum-\emph{capacity-weighted} priced link before \textit{ROUND} $m$. Since the \emph{capacity-weighted} price of this link  among all the \textit{OPEN} attached links is the highest, it is also the local maximum among the \textit{OPEN} attached links received from the $(K+1)$-hop neighbourhood. Thus link $l^m$ gets \textit{MARKED}. Since a \textit{MARKED} link will always remain in the same state, 
$$\textrm{if } l \in \mathcal{L}^m_M, \textrm{ then } l \in \mathcal{L}^k_M, \forall k \geq m+1 \Longrightarrow \mathcal{L}^m_M  \subseteq \mathcal{L}^{m+1}_M $$
Now let us consider link $l^m$, 
$$ l^m \in \mathcal{L}^m_O \Longrightarrow l^m \notin \mathcal{L}^m_M$$
But from the previous argument, link $l^m$ gets \textit{MARKED} in \textit{ROUND} $m+1$. Thus
\begin{equation} \label{eqn:finite_time_eq1}
l^m \in \mathcal{L}^{m+1}_M\ and\ l^m \notin \mathcal{L}^m_M \Longrightarrow \mathcal{L}^m_M  \subset \mathcal{L}^{m+1}_M 
\end{equation}
Since a link \textit{CLOSED} in \textit{ROUND} $m$, will remain \textit{CLOSED} for the subsequent \textit{ROUND}s, we have
\begin{equation} \label{eqn:finite_time_eq2}
\textrm{if } l \in \mathcal{L}^m_C, \textrm{ then } l \in \mathcal{L}^k_C, \forall k \geq m+1 \Longrightarrow \mathcal{L}^m_C  \subseteq \mathcal{L}^{m+1}_C 
\end{equation}
Now from \eqref{eqn:finite_time_eq1} and \eqref{eqn:finite_time_eq2} we have
$$ \mathcal{L}^m_C \cup \mathcal{L}^m_M \subset \mathcal{L}^{m+1}_C \cup \mathcal{L}^{m+1}_M $$ 
At all times, a link $l$ can be in one of the four states, i.e.,
$$ \forall m,\mathcal{L}^m_C \cup \mathcal{L}^m_M \cup \mathcal{L}^m_O \cup \mathcal{L}^m_H = \mathcal{L}$$
From the above two argument,
$$ \mathcal{L}^{m+1}_O \cup \mathcal{L}^{m+1}_H \subset \mathcal{L}^{m}_O \cup \mathcal{L}^{m}_H $$
Since the number of links in set $\mathcal{L}$ is finite, there exists a $t< \infty$, such that
$$ \mathcal{L}^{t}_O \cup \mathcal{L}^{t}_H = \lbrace \phi \rbrace $$
Thus the algorithm terminates in finite number of \textit{ROUND}s and thus in finite time. \qed
\end{proof}

\subsection*{\textbf{Proof of Lemma \ref{lemma:ft_lemma2}}}
\begin{proof}
Assume that there is no such $j \in \mathcal{L}^m_M$ for some $i \in \mathcal{L}^m_C$.
Then the link $i$ would not have received any \textit{MARKED} link that interferes with it, in slot $S^{m-1}_M$ (Algorithm 2, At time $T^m_M$, lines 5, 6). Then, this would imply that either link $i$ would be \textit{OPEN}ed or would be in \textit{CHECK}. i.e.,
$$ i \in \mathcal{L}^m_H \cup \mathcal{L}^m_O \Longrightarrow i \notin \mathcal{L}^m_C $$
But this is a contradiction. Thus there exists a link $j \in \mathcal{L}^m_M$, such that $d(i,j) < K$. \qed
\end{proof}

\subsection*{\textbf{Proof of Lemma \ref{lemma:ft_lemma3}}}
\begin{proof}
Let us assume that there is no $j \in \mathcal{L}^m_O$ for some $i \in \mathcal{L}^m_H$, such that $\alpha_i p_i < \alpha_j p_j$. This would imply that $\alpha_i p_i > \alpha_j p_j, \forall j \in \mathcal{L}^m_O$. Since link $i \in \mathcal{L}^m_H$, we can say that a link $k$, such that $\alpha_k p_k > \alpha_i p_i$ and $d(k,i)=0$ was \textit{OPEN}ed at time $T^{m-1}_M$ (Algorithm 2, At time $T^m_M$, line 8). Since link $k$ was \textit{OPEN}ed at time $T^{m-1}_M$, $k \in \mathcal{L}^m_O$. Let
$$\alpha_i p_i > \alpha_j p_j, \forall j \in \mathcal{L}^m_O$$
But we have $\alpha_k p_k > \alpha_i p_i$, Thus
$$\alpha_k p_k > \alpha_j p_j, \forall j \in \mathcal{L}^m_O \Longrightarrow k \notin \mathcal{L}^m_O$$
But this is a contradiction to the statement that $k \in \mathcal{L}^m_O$. Thus $\forall i \in \mathcal{L}^m_H, \alpha_i p_i < \alpha_j p_j,\textrm{for some } j \in \mathcal{L}^m_O$. Let
$$ \alpha p^m_{max} = \max_{l \in \mathcal{L}, l \notin \mathcal{L}^m_M \cup \mathcal{L}^m_C}{\alpha_l p_l} = \max_{l \in \mathcal{L}^m_O \cup \mathcal{L}^m_H}{\alpha_l p_l} = \max{(\max_{k \in \mathcal{L}^m_O}{\alpha_k p_k},\max_{l \in \mathcal{L}^m_H}{\alpha_l p_l})}$$
Let
$$ i = arg \max_{l \in \mathcal{L}^m_H}{\alpha_l p_l}$$
It is evident that $i \in \mathcal{L}^m_H$. Thus from the previous claim, there exists a $j \in \mathcal{L}^m_O$ such that
$$\alpha_j p_j > \alpha_i p_i$$
i.e.,
$$\alpha_j p_j > \max_{l \in \mathcal{L}^m_H}{\alpha_l p_l}$$
Also
$$ \max_{k \in \mathcal{L}^m_O} \alpha_k p_k \geq \alpha_j p_j > \max_{l \in \mathcal{L}^m_H}{\alpha_l p_l}$$
Hence
$$\alpha p^m_{max} = \max_{k \in \mathcal{L}^m_O}{\alpha_k p_k}$$
Thus $\arg \, \alpha p^m_{max} \in \mathcal{L}^m_O $. \qed
\end{proof}

\subsubsection*{\textbf{Proof of Theorem \ref{theorem:same_links}}}
\begin{proof} 
Let $\mathcal{L}_C$ be the set of links \textit{CHOSEN} by the centralized greedy algorithm. \\
Let the set $\mathcal{L}_C$ be ordered and indexed in the decreasing order of link \emph{capacity-weighted} price as $\lbrace l_1, l_2,...,l_v... \rbrace$. \\
Let the distributed greedy algorithm terminate after $t$ \textit{ROUND}s. Let $\mathcal{L}^{t+1}_M$ be the set of \textit{MARKED} links after the termination of the algorithm. 

We need to prove that every link \textit{CHOSEN} by the centralized greedy algorithm is \textit{MARKED} by the distributed greedy algorithm, by the time it terminates i.e., $\mathcal{L}_C \subseteq  \mathcal{L}^{t+1}_M$. We will prove the above claim via induction. \\

\noindent
\textit{Induction statement:} If links $l_1,l_2,...,l_k \in \mathcal{L}_C$ then $l_1,l_2,...,l_k \in \mathcal{L}^{t+1}_M$. \\
\textit{Basis:} To show the statement holds for the globally maximum \emph{capacity-weighted} priced link.\\

Let link $l_1 \in \mathcal{L}_C$ be the globally maximum \emph{capacity-weighted} priced link. Thus, this link will also be a local maximum among interfering links in a $(K + 1)$-hop neighbourhood. Hence, this link will be \textit{MARKED} after the $1^{st}\ ROUND$. i.e.,  $l_1 \in \mathcal{L}^{2}_M$. Since \textit{MARKED} is an absorbing state, $ l_1 \in \mathcal{L}^{t+1}_M$.
Now, let us define 
$$\mathcal{I}(y) = \lbrace l \in  \mathcal{L}:d(l, y) < K \rbrace$$
as the set of links that interfere with link $y$. Let 
$$ \mathcal{L}^{k+1} = \mathcal{L} - \cup^k_{i=1}(l_i \cup \mathcal{I}(l_i))$$
be the set of links left after links $\lbrace l_1,l_2,...l_k \rbrace$ are \textit{CHOSEN}. Let
\begin{equation} \label{eqn:th_same_set_eq1}
\forall l \in \mathcal{L}^{k+1}, \mathcal{P}(l) = \lbrace l^{'} \in \mathcal{L}^{k+1}:d(l,l^{'}) < K, \alpha_{l^{'}} p_{l^{'}} > \alpha_l p_{l} \rbrace
\end{equation}
It is obvious that for link $L_{k+1}$ to be \textit{CHOSEN}, $\mathcal{P}(l_{k+1})= \lbrace \phi \rbrace$ .  \\

\noindent
\textit{Inductive step:} If  $l_1,l_2,...,l_k \in \mathcal{L}^{t+1}_M$ given that  $l_1,l_2,...,l_k \in \mathcal{L}_C$, then if $l_{k+1} \in \mathcal{L}_C$ then $l_{k+1} \in \mathcal{L}^{t+1}_M$. 
Since $l_1,l_2,...,l_k \in \mathcal{L}^{t+1}_M$, for each $i \in \lbrace 1,2,...,k \rbrace$
$$ \exists m_i \leq t+1 : l_i \in \mathcal{L}^{m_i}_M, l_i \notin \mathcal{L}^{s}_{M} \textrm{ for } s < m_i$$
Since \textit{MARKED} is an absorbing state,
$$l_i \in \mathcal{L}^{r}_M, \quad  \forall m_i \leq r \leq t+1 \quad  $$
Let
$$\forall l \in \mathcal{L}^{m}_O, \mathcal{P}^{m}(l) = \lbrace l^{'} \in \mathcal{L}^m_O : d(l,l^{'}) < K, \alpha_{l^{'}} p_{l^{'}} > \alpha_l p_{l} \rbrace$$
We note that link $l$ is \textit{MARKED} in \textit{ROUND} $m$, if $\mathcal{P}^{m}(l)= \lbrace \phi \rbrace $. Let
$$ m^{'} = \max^{k}_{i=1}{m_i}$$
It is easy to see that before the end of \textit{ROUND} $m^{'}$, links $\lbrace l_1,l_2,...l_k \rbrace$ are \textit{MARKED} and the links that interfere with these links are \textit{CLOSED}. Thus
\begin{equation} \label{eqn:th_same_set_eq2}
\mathcal{L}^{m^{'}}_O \subseteq \mathcal{L} - \cup^k_{i=1}(l_i \cup \mathcal{I}(l_i)) \subseteq \mathcal{L}^{k+1} 
\end{equation}
Now,
$$ \mathcal{P}^{m^{'}}(l_{k+1}) = \lbrace l^{'} \in \mathcal{L}^{m^{'}}_O : d(l_{k+1},l^{'}) < K, \alpha_{l^{'}} p_{l^{'}} > \alpha_{l_{k+1}} p_{l_{k+1}} \rbrace \quad \forall \, \l^{k+1} \in \mathcal{L}^{m^{'}}_{O}$$
From \eqref{eqn:th_same_set_eq1} and \eqref{eqn:th_same_set_eq2}, we can say that
$$\mathcal{P}^{m^{'}}(l_{k+1}) \subseteq \mathcal{P}(l_{k+1}) $$
Since $l_{k+1} \in \mathcal{L}_C,\ \mathcal{P}(l_{k+1})= \lbrace \phi \rbrace $. Thus
$$ \mathcal{P}^{m^{'}}(l_{k+1}) \subseteq \lbrace \phi \rbrace \Longrightarrow  \mathcal{P}^{m^{'}}(l_{k+1}) = \lbrace \phi \rbrace  $$
Now, since the algorithm terminates after $t$ \textit{ROUND}s, 
$$\mathcal{L}^{t+1}_O \cup \mathcal{L}^{t+1}_H = \lbrace \phi \rbrace $$
$$\therefore l_{k+1} \notin \mathcal{L}^{t+1}_O \Longrightarrow m^{'} \neq t+1$$
Thus link $l_{k+1}$ gets \textit{MARKED} in \textit{ROUND} $m^{'} \leq t$. Thus link $l_{k+1}$ gets \textit{MARKED} before the algorithm terminates. Therefore, 
$$\forall l \in \mathcal{L}_C,\ l \in \mathcal{L}^{t+1}_M \Longrightarrow  \mathcal{L}_C \subseteq \mathcal{L}^{t+1}_M$$
Now, let us assume that LHS is a strict subset of RHS, i.e.,
$$ \mathcal{L}_C \subset \mathcal{L}^{t+1}_M $$
Then there exists a link $l_i$ such that $l_i \notin \mathcal{L}_C$ but $l_i \in \mathcal{L}^{t+1}_M$. Since $l_i \in \mathcal{L}^{t+1}_M$, we can say that
$$d(l,l_i) \geq K, \forall l \in \mathcal{L}^{t+1}_M$$
Since $ \mathcal{L}_C \subset \mathcal{L}^{t+1}_M $, 
$$d(l,l_i) \geq K, \forall l \in \mathcal{L}_C$$
If the above was true, then $l_i \in \mathcal{L}_C$. But this contradicts our assumption that $ \mathcal{L}_C \subset \mathcal{L}^{t+1}_M $. Thus LHS can not be a strict subset of RHS.
$$\Longrightarrow  \mathcal{L}_C = \mathcal{L}^{t+1}_M$$ \qed
\end{proof}

\section*{Appendix B}

\subsection*{\textbf{Proof of Proposition \ref{prop:subgradient}}}
\begin{proof}
Let $\mathbf{p}_1, \mathbf{p}_2 \in  \mathcal{R}^L_{+}$. Let $x_f(\mathbf{p})$ and $y_f(\mathbf{p})$ represent the optimal flow rate and optimal routing
vector of a flow $f \in \mathcal{F}$ respectively. $\mathbf{y}(\mathbf{p})$ represents a $L$ sized column vector, $l^{\mathrm{th}}$ entry of which
indicates the aggregate traffic of all the flows carried by the link $l$ for the price vector $p$, i.e., $\sum_{f \in \mathcal{F}} y_f(\mathbf{p})$. Then at $\mathbf{p}_2$ we have the following:
\begin{eqnarray}
D(\mathbf{p}_2) = & \sum_{f\in \mathcal{F}} U(x_f(\mathbf{p}_2)) - \mathbf{p}^{T}_{2}(\mathbf{y}(\mathbf{p}_2) - \mathbf{c}_{opt}(\mathbf{p}_2)) \nonumber \\
\geq & \sum_{f\in \mathcal{F}} U(x_f(\mathbf{p}_1)) - \mathbf{p}^{T}_{2}(\mathbf{y}(\mathbf{p}_1) - \mathbf{c}_{dgrd}(\mathbf{p}_1)) \nonumber \\
\geq & \sum_{f\in \mathcal{F}} U(x_f(\mathbf{p}_1)) - \mathbf{p}^{T}_{1}(\mathbf{y}(\mathbf{p}_1) - \mathbf{c}_{opt}(\mathbf{p}_1)) \nonumber \\
& - (\mathbf{p}_{2} - \mathbf{p}_{1})^{T}(\mathbf{y}(\mathbf{p}_1) - \mathbf{c}_{dgrd}(\mathbf{p}_1)) \nonumber \\
& - \mathbf{p}^{T}_{1}(\mathbf{c}_{opt}(\mathbf{p}_1) - \mathbf{c}_{dgrd}(\mathbf{p}_1)) \nonumber \\
\geq & D(\mathbf{p}_1) + (\mathbf{p}_{2} - \mathbf{p}_{1})^{T}(\mathbf{c}_{dgrd}(\mathbf{p}_1)-\mathbf{y}(\mathbf{p}_1)) \label{eq:aeq1} \\
& -  \mathbf{p}_{1}^{T}(\mathbf{c}_{opt}(\mathbf{p}_1) - \mathbf{c}_{dgrd}(\mathbf{p}_1)) \nonumber
\end{eqnarray}
From the definition of the $\epsilon$-subgradient, we have
\begin{eqnarray} \label{eq:aeq2}
D(\mathbf{p}_{2}) \geq D(\mathbf{p}_{1}) - \epsilon + (\mathbf{p}_{2} - \mathbf{p}_{1})^T h(\mathbf{p}_{1}) 
\end{eqnarray}
By comparing Equation \ref{eq:aeq1} and Equation \ref{eq:aeq2}, we can write 
$$h(\mathbf{p}) = \mathbf{c}_{dgrd}(\mathbf{p}) - \mathbf{y}(\mathbf{p})$$
Also note that $\mathbf{p}_{1}^{T}(\mathbf{c}_{dgrd}(\mathbf{p}_1) - \mathbf{c}_{opt}(\mathbf{p}_1))$ is always a positive quantity, thus we can write
$$\epsilon(\mathbf{p}) = \mathbf{p}^{T}(\mathbf{c}_{opt}(\mathbf{p}) - \mathbf{c}_{dgrd}(\mathbf{p})) $$ \qed
\end{proof}

\subsection*{\textbf{Proof of Theorem \ref{th:convergence}}}
\begin{proof} 
Let $\mathbf{p}^{*}$ be the optimal price vector of the dual problem. From the price update equation, we have
$$\mathbf{p}[j+1] = (\mathbf{p}[j] + \delta(\mathbf{y}[j] - \mathbf{c}[j]))^{+} $$
Equivalently,
\begin{eqnarray}
||\mathbf{p}[j+1] - \mathbf{p}^{*}||^{2}_{2} &\leq & || \mathbf{p}[j] + \delta(\mathbf{y}[j] - \mathbf{c}[j])  - \mathbf{p}^{*}||^{2}_{2} \nonumber \\
&=& || \mathbf{p}[j] - \mathbf{p}^{*}||^{2}_{2} + \delta^2|| \mathbf{y}[j] - \mathbf{c}[j]||^{2}_{2} + \nonumber \\
&& 2\delta (\mathbf{p}[j] - \mathbf{p}^{*})^{T} (\mathbf{y}[j] - \mathbf{c}[j]) \label{eq:a_eq1}
\end{eqnarray}
From the definition of an $\epsilon$-subgradient, we have
$$D(\overline{\mathbf{p}}) \geq D(\mathbf{p}) - \epsilon + (\overline{\mathbf{p}} - \mathbf{p})^T \mathbf{h}(\mathbf{p}), \, \,  \forall \, \overline{\mathbf{p}} \in \mathcal{R}^n$$
Hence, we have
\begin{eqnarray}
(\mathbf{p}[j] - \mathbf{p}^{*})^{T} (\mathbf{y}[j] - \mathbf{c}[j]) \leq -(D(\mathbf{p}[j]) - D(\mathbf{p}^{*}) + \epsilon(\mathbf{p}[j])) \label{eq:a_eq2}
\end{eqnarray}
Henceforth, we represent $\epsilon(\mathbf{p}[j])$ as $\epsilon_j$ for convenience. Substituting Equation \eqref{eq:a_eq2} in Equation \eqref{eq:a_eq1}, we get
$$||\mathbf{p}[j+1] - \mathbf{p}^{*}||^{2}_{2} \leq || \mathbf{p}[j] - \mathbf{p}^{*}||^{2}_{2} + \delta^2|| \mathbf{y}[j] - \mathbf{c}[j]||^{2}_{2} - 2\delta (D(\mathbf{p}[j]) - D(\mathbf{p}^{*}) + \epsilon_j) $$
Applying the inequalities recursively, we obtain
\begin{eqnarray*}
||\mathbf{p}[j+1] - \mathbf{p}^{*}||^{2}_{2} \leq & || \mathbf{p}[1] - \mathbf{p}^{*}||^{2}_{2} + \delta^2 \sum^{j}_{\tau = 1 }|| \mathbf{y}[\tau] - \mathbf{c}[\tau]||^{2}_{2} \\
& - 2\delta (\sum^{j}_{\tau = 1 }(D(\mathbf{p}[\tau]) - D(\mathbf{p}^{*})) + \sum^{j}_{\tau = 1 } \epsilon_{\tau}) 
\end{eqnarray*}
Since $||\mathbf{p}[j+1] - \mathbf{p}^{*}||^{2}_{2} \geq 0, \, \forall j$, we get
$$ 2\delta \sum^{j}_{\tau = 1 }(D(\mathbf{p}[\tau]) - D(\mathbf{p}^{*})) \leq || \mathbf{p}[1] - \mathbf{p}^{*}||^{2}_{2} + \delta^2\sum^{j}_{\tau = 1 }|| \mathbf{y}[\tau] - \mathbf{c}[\tau]||^{2}_{2} + \sum^{j}_{\tau = 1 } \epsilon_{\tau} $$

Since we assume that $|| \mathbf{y}[\tau] - \mathbf{c}[\tau] ||_2 = ||\mathbf{h}(\mathbf{p}[\tau]) || \leq H$, we have 
$$ \lim_{j \to \infty} \sup \frac{1}{j} \sum^{j-1}_{\tau = 0 }(D(\mathbf{p}[\tau]) - D(\mathbf{p}^{*})) \leq \frac{\delta H^2}{2} + \lim_{j \to \infty} \sup \frac{1}{j}\sum^{j-1}_{\tau = 0 } \epsilon_{\tau} $$

Since we assume that $\lim_{j \to \infty} \epsilon_j \leq \epsilon$, there exists $j_0$ such that $\epsilon_j \leq \epsilon , \, \forall j \geq j_0$. Therefore, for large enough $j$, we have 
$$\limsup_{j \to \infty} \frac{1}{j} \sum^{j}_{\tau = 1 } \epsilon_{\tau} \leq \limsup_{j \to \infty} \frac{1}{j}\sum^{j_0}_{\tau = 1 } \epsilon_{\tau} + \limsup_{j \to \infty} \frac{(j - j_0)}{j} \epsilon = \epsilon$$ \qed
\end{proof}
	
\end{document}